\documentclass[nofootinbib,twocolumn,preprintnumbers]{revtex4}
\usepackage{amsmath}
\usepackage{amssymb}
\usepackage{graphicx}
\usepackage{booktabs}

\newcommand{\GeV}{\;\mathrm{GeV}}

\newcommand{\jet}{\text{jet}}
\begin{document}

\title{Pileup subtraction for jet shapes}

\preprint{CERN-PH-TH/2012-300}
\preprint{LPN12-120}

\author{Gregory Soyez$^{1}$, Gavin P. Salam$^{2,3,4}$, Jihun Kim$^{1}$, 
Souvik Dutta$^{5}$ and Matteo Cacciari$^{3,6}$\vspace{1em}}

\affiliation{
\mbox{$^1$IPhT, CEA Saclay, CNRS URA 2306, F-91191 Gif-sur-Yvette, France}\\
$^2$CERN, PH-TH, CH-1211 Geneva 23, Switzerland\\
$^3$LPTHE; CNRS UMR 7589; UPMC Univ.\ Paris 6; Paris 75252, France\\
\mbox{$^4$Department of Physics, Princeton University, Princeton, NJ 08544, USA}\\
$^5$IIT Bombay, Mumbai, MH 400076, India\\
$^6$Universit\'e Paris Diderot, Paris, France
}

\begin{abstract}
  Jet shapes have the potential to play a role in many LHC analyses,
  for example in quark-gluon discrimination or jet substructure
  analyses for hadronic decays of boosted heavy objects.
  Most shapes, however, are significantly affected by pileup.
  We introduce a general method to correct for pileup effects in
  shapes, which acts event-by-event and jet-by-jet, and accounts also
  for hadron masses.
  It involves a numerical determination, for each jet, 
  of a given shape's susceptibility to pileup.
  Together with existing techniques for determining the level of
  pileup, this then enables an extrapolation to zero pileup.
  The method can be used for a wide range of jet shapes and we show
  its successful application in the context of quark/gluon discrimination
  and top-tagging.
\end{abstract}


\maketitle

When energetic quarks or gluons (partons) fragment, they produce
collimated bunches of hadrons known as jets.
Jets mostly conserve the energy and direction of the originating
parton,
consequently they have long been used at colliders as a stand-in for
generic partons, as is the case currently at CERN's Large Hadron
Collider (LHC).
In recent years extensive interest has developed in going
beyond this basic use: for example, to understand if the parton is a
quark or a gluon, or to identify rare cases where a single jet
originated from multiple hard partons, perhaps from the hadronic decay
of a highly-boosted W, Z or Higgs boson, top quark or other massive
object~\cite{Abdesselam:2010pt,Altheimer:2012mn,Plehn:2011tg}.
The ``jet substructure'' techniques being developed in this context
will be crucial to exploit the full kinematic reach of the LHC,
notably the high transverse-momentum (high-$p_t$) region, and to
maximise the LHC's sensitivity to hadronic manifestations of new
physics scenarios.

Two key classes of approach are available to probe substructure: one
identifies smaller ``subjets'' within a larger jet and then perform
selections based on the kinematics of those subjets, for
example~\cite{Seymour:1993mx,Butterworth:2002tt,Butterworth:2008iy,Kaplan:2008ie,Ellis:2009me,Krohn:2009th,Plehn:2009rk};
the other involves jet-shape observables, sensitive
to the geometrical spread of the energy within the jet,
e.g.~\cite{Thaler:2008ju,Almeida:2008yp,Kim:2010uj,Thaler:2010tr,Thaler:2011gf,Hook:2011cq,Gallicchio:2011xq,Feige:2012vc}.
Both classes appear to be powerful and viable experimentally (see
e.g.~\cite{ShapeMeasurements,Chatrchyan:2012ku}) and ultimate
performance in exploiting jet substructure will probably be obtained
through some combination of them.

One potential show-stopper in substructure studies is the problem of
pileup: with the LHC now operating at high instantaneous luminosities,
each interesting, high-$p_t$ proton-proton collision is accompanied by
dozens of additional $pp$ collisions, which add substantial low-$p_t$
noise to the event. 
Pileup modifies a jet's kinematics, on average shifting its $p_t$
in proportion to the level of noise in the event and to the jet's
extent, or ``area''~\cite{Cacciari:2008gn}, in rapidity ($y = \frac12 \ln
\frac{E+p_z}{E-p_z}$) and azimuth ($\phi$).
Two techniques are in common use to correct for this: the removal of
an ``offset'' from the jet in proportion to the number of observed
pileup events~\cite{ATLASPileup}; and the ``area--median'' method,
which subtracts an amount given by the product of the event's measured
pileup $p_t$ density ($\rho$) and the jet's measured area
($A$)~\cite{Cacciari:2007fd,ATLASJetArea,CMSPileup}.\footnote{
  With particle flow~\cite{ParticleFlow}, one can also
  directly discard the charged component of
  pileup~\cite{CMSPileup}. The remaining neutral part
  must be subtracted in some other way (it may not be the expected fraction of the charged
  component, due to detector effects).}
While the second of these methods can be straightforwardly applied also to
subjets, jet shapes have so far proved more challenging to correct.

Jet shapes are particularly sensitive to pileup because its diffuse
soft energy flow is characteristically different from the more
collimated distribution of energy due to normal jet
fragmentation.
One can attempt to mitigate pileup's impact by determining the shape
using just charged tracks, or by breaking a jet into subjets and using
just the hardest subjets; but both methods throw away a
significant fraction of the original particles contributing to the
jet's shape, introducing a bias.
One can also carry out analytical calculations of a given shape's
sensitivity, as in Refs.~\cite{Sapeta:2010uk,Alon:2011xb}, or add in
particles from a ``complementary'' cone at 90 degrees to the jet's axis
in order to determine an average sensitivity~\cite{Alon:2011xb}.
These methods have so far, however, been limited either to specific
observables, restricted classes of jets (e.g.\ circular jets), or low
pileup.
The intent of this letter is to develop an effective, simple, general
method to correct jet shapes for pileup.

Our approach is related to the area--median
method, which has been found to be beneficial in both
ATLAS~\cite{ATLASJetArea} and CMS~\cite{CMSPileup} (see
also~\cite{AlcarazMaestre:2012vp}). 
It is intended to be valid for arbitrary jet algorithms and generic
infrared and collinear safe jet shapes,\footnote{For the correction of
  collinear unsafe quantities, e.g.\ fragmentation function moments,
  as used for quark/gluon discrimination
  in~\cite{Chatrchyan:2012sn},
  see~\cite{Cacciari:2012mu}.} without the need for 
dedicated analytic study of each individual shape variable.
It also involves an extension of the original area--median
prescription to account for hadron masses.

The first ingredient is a characterisation of the average pileup
density in a given event in terms of two variables, $\rho$ and
$\rho_m$, such that the 4-vector of the expected pileup deposition in
a small region of size $\delta y \delta \phi$ can be written
\begin{equation}
  \label{eq:rho-rhom-parametrisation}
  \left[\,
    \rho \cos\phi, \,
    \rho \sin\phi, \,
    (\rho + \rho_m) \sinh y,\,
    (\rho + \rho_m) \cosh y \,
  \right] \delta y \delta\phi\,,
\end{equation}
where $\rho$ and $\rho_m$ have only weak dependence on $y$ (and $\phi$).
Relative to the original area--median proposal~\cite{Cacciari:2007fd},
a novelty here is the inclusion of a term $\rho_m$.
It arises because pileup consists of low-$p_t$ hadrons, and
their masses are not negligible relative to their $p_t$ (cf.\
also~\cite{Salam:2001bd,Mateu:2012nk}).
It is important mainly for observables sensitive to differences
between energy and 3-momentum, e.g.\ jet masses, as we will see below.

The second and main new ingredient is a determination, for a specific jet, of the
shape's sensitivity to pileup.
Let the shape be defined by some function $V(\{p_i\}_\text{jet})$ of
the momenta $p_i$ in the jet.
Among these momenta, we include a set of
``ghosts''~\cite{Cacciari:2008gn}, very low momentum particles that
cover the $y-\phi$ plane at high density, each of them mimicking a
pileup-like component in a region of area $A_g$.
%
We then consider the derivatives of the jet shape
with respect to the transverse momentum scale, $p_{t,g}$, of the
ghosts and with respect to a component $m_{\delta,g} \equiv
\sqrt{\smash[b]{m^2_g+p_{t,g}^2}} - p_{t,g}$,
\begin{equation}
  \label{eq:Vnm}
  V^{(n,m)}_\text{jet} \equiv A_g^{n+m}\,
    \partial_{p_{t,g}}^n \, \partial_{m_{\delta,g}}^m \, V(\{p_i\}_\text{jet})\,.
\end{equation}
The derivatives are to be evaluated at $p_{t,g} = m_{\delta,g} = 0$,
and by scaling all ghost momenta simultaneously.

Given the level of pileup, $\rho, \rho_m$, and the information on the
derivatives, one can then extrapolate the value of the jet's shape to
zero pileup,
\begin{multline}
  \label{eq:shape-subtraction-with-rhom}
  V_\text{jet,sub} = V_\text{jet} - \rho V^{(1,0)}_\text{jet}
  - \rho_m V^{(0,1)}_\text{jet}\\
  + \frac12 \rho^2 V^{(2,0)}_\text{jet} 
  + \frac12 \rho_m^2 V^{(0,2)}_\text{jet} 
  +  \rho \rho_m V^{(1,1)}_\text{jet} 
  + \cdots\,.
\end{multline}
where the formula takes into account the fact that the derivatives
are evaluated for the jet including the pileup.

Handling derivatives with respect to both $p_{t,g}$ and $m_{\delta,g}$ can
be cumbersome in practice. 
An alternative is to introduce a new variable $r_{t,g}$ and set
$p_{t,g} = r_{t,g}$ and $m_{\delta,g} = \frac{\rho_m}{\rho} r_{t,g}$.
We then take total derivatives with respect to $r_{t,g}$
\begin{equation}
  \label{eq:Vntotal}
  V^{[n]}_\text{jet} \equiv A_g^n \, \frac{d^n}{dr_{t,g}^n}
  V(\{p_i\}_\text{jet})\,,
\end{equation}
so that the correction can be rewritten
\begin{equation}
  \label{eq:shape-subtraction-with-rho-Vntotal}
  V_\text{jet,sub} = V_\text{jet} - \rho V^{[1]}_\text{jet}
   + \frac12 \rho^2 V^{[2]}_\text{jet} + \cdots\,.
\end{equation}

The derivatives $V^{(m,n)}$ or $V^{[n]}_\text{jet}$ can be determined
numerically, for a specific jet, by rescaling the ghost momenta and
reevaluating the jet shape for multiple rescaled values.
Typically this is more stable with
Eq.~(\ref{eq:Vntotal}) and this is the
approach we use below.

To investigate the performance of our correction procedure, we
consider a number of jet shapes:
\begin{itemize}
\item Angularities~\cite{Berger:2003iw,Almeida:2008yp}, adapted to
  hadron-collider jets as $\theta^{(\beta)} = \sum_i p_{ti} \Delta
  R_{i,\jet}^{\beta}/ \sum_i p_{ti}$, for $\beta = 0.5,1,2,3$;
  $\theta^{(1)}$, the ``girth'', ``width'' or ``broadening'' of the
  jet, has been found to be particularly useful for quark/gluon
  discrimination~\cite{Gallicchio:2011xq,ATLASqgdiscr}.
\item Energy-energy-correlation (EEC) moments, advocated for their
  resummation simplicity in~\cite{Banfi:2004yd}, $E^{(\beta)} =
  \sum_{i,j} p_{ti} p_{tj} \Delta R_{i,j}^{\beta}/ 
  (\sum_i p_{ti})^2$, using the same set of $\beta$
  values. EEC-related variables have been studied recently also
  in~\cite{Jankowiak:2011qa}. 
\item ``Subjettiness'' ratios, designed for characterising
  multi-pronged jets~\cite{Kim:2010uj,Thaler:2010tr,Thaler:2011gf}:
  one defines the subjettiness $\tau_N^{(\text{axes},\beta)} = $
  $\sum_{i} p_{ti} \min(\Delta R_{i1},\ldots,\Delta
  R_{iN})^\beta/\sum_i p_{ti}$, where $\Delta R_{ia}$ is the distance
  between particle $i$ and axis $a$, where $a$ runs from $1$ to $N$.
  %
  One typically considers ratios such as $\tau_{21} \equiv
  \tau_2/\tau_1$ and $\tau_{32} \equiv \tau_3/\tau_2$ (the latter used
  e.g.\ in a recent search for R-parity violating gluino
  decays~\cite{ATLAS:2012dp}); we consider $\beta=1$ and $\beta=2$, as
  well as two choices for determining the axes: 
  ``$\text{kt}$'', which exploits the $k_t$
  algorithm~\cite{Catani:1993hr,Kt-EllisSoper} to decluster the jet to
  $N$ subjets and then uses their axes;
  and ``$\text{1kt}$'', which adjusts the ``$\text{kt}$'' axes so as
  to obtain a single-pass approximate minimisation of
  $\tau_N$~\cite{Thaler:2011gf}.
%
%

\item A longitudinally invariant version of the planar
  flow~\cite{Thaler:2008ju,Almeida:2008yp}, involving a $2\times2$
  matrix $M_{\alpha\beta} = \sum_{i} p_{ti} (\alpha_i -
  \alpha_{\text{jet}})(\beta_i - \beta_{\text{jet}})$, where $\alpha$
  and $\beta$ correspond either to the rapidity $y$ or azimuth
  $\phi$; the planar flow is then given by $\text{Pf} =
  4\lambda_1\lambda_2/(\lambda_1 + \lambda_2)^2$, where
  $\lambda_{1,2}$ are the two eigenvalues of the matrix.
\end{itemize}
\begin{figure*}[th]
  %
  \includegraphics[angle=270,width=0.32\textwidth]{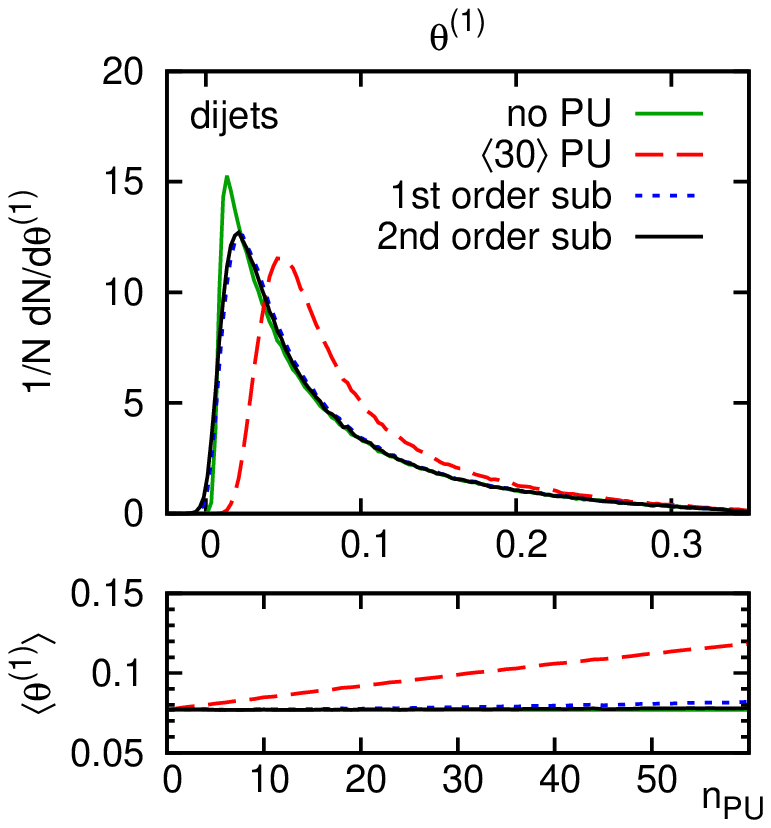}\hfill
  \includegraphics[angle=270,width=0.32\textwidth]{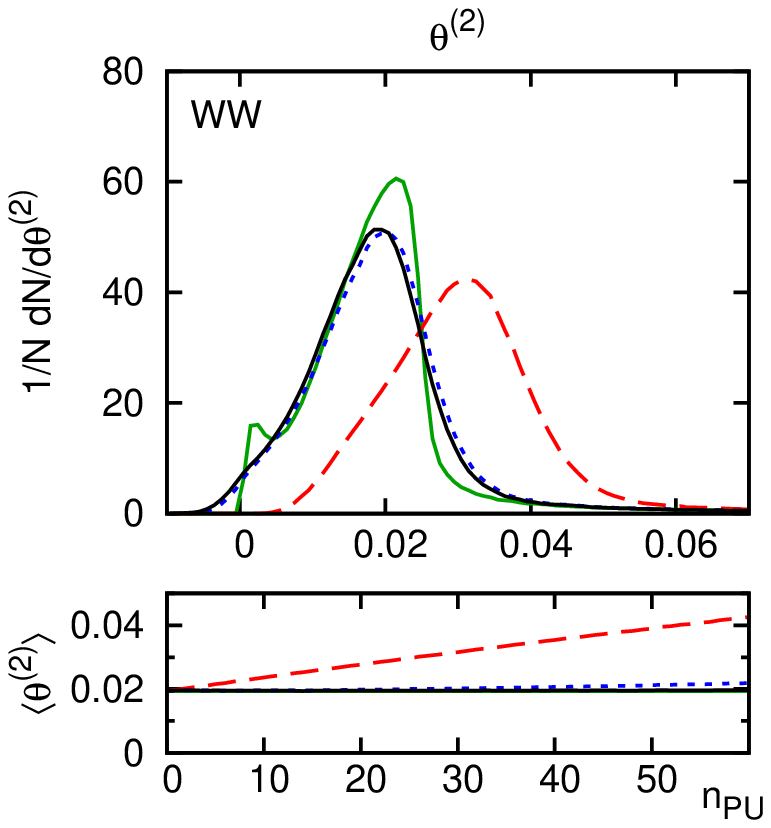}\hfill
  \includegraphics[angle=270,width=0.32\textwidth]{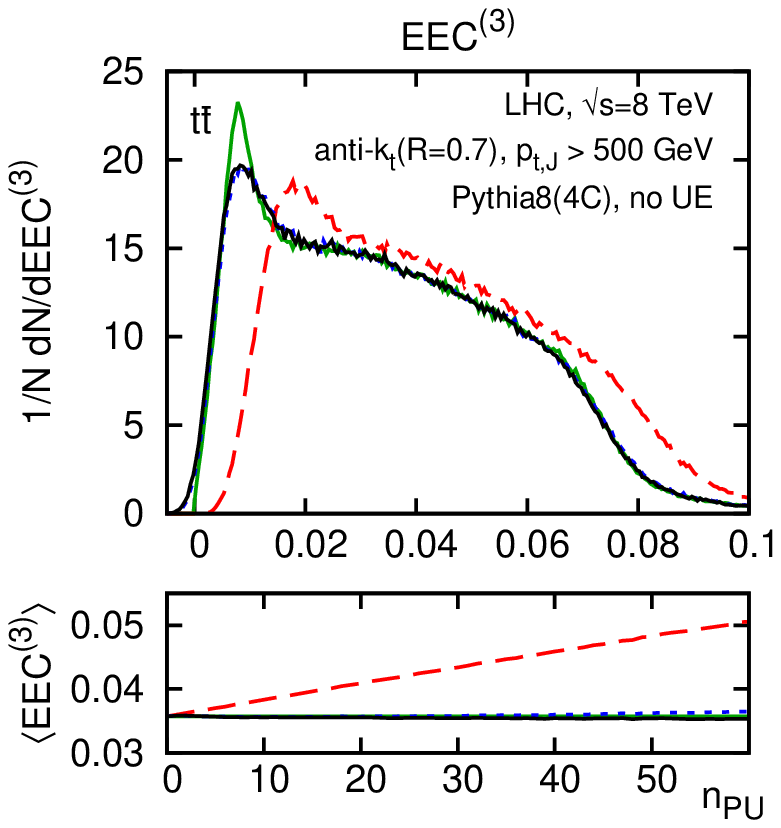}\hfill
  \includegraphics[angle=270,width=0.32\textwidth]{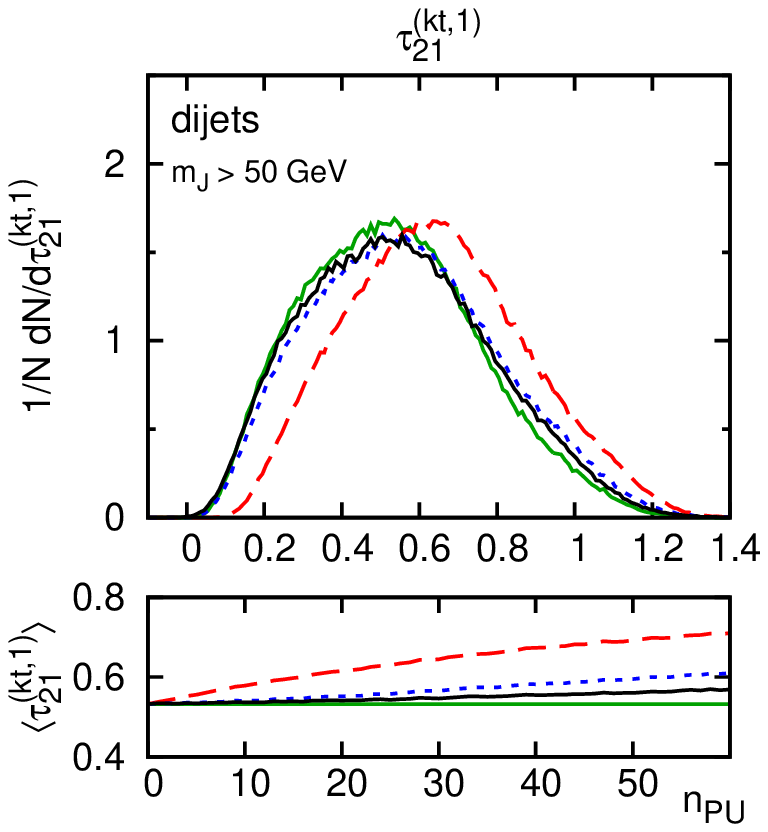}\hfill
  \includegraphics[angle=270,width=0.32\textwidth]{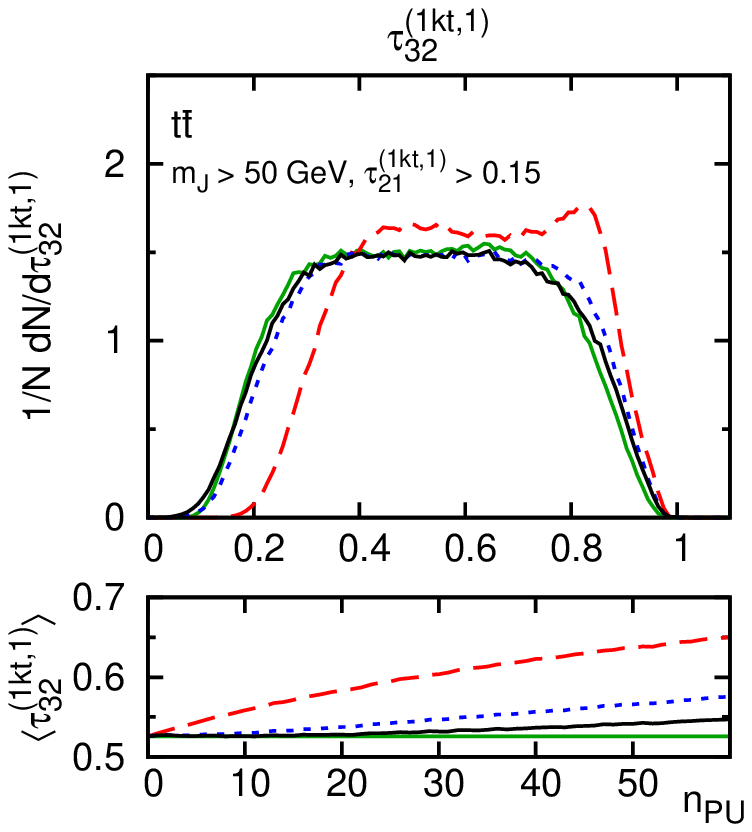}\hfill
  \includegraphics[angle=270,width=0.32\textwidth]{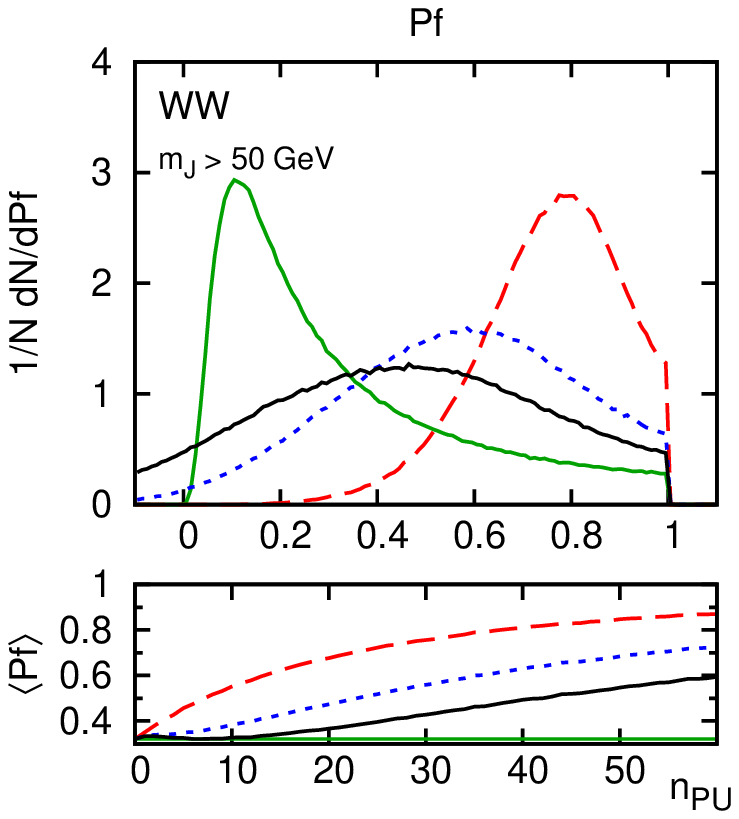}
  \centering
  \caption{Impact of pileup and subtraction on various jet-shape
    distributions and their averages, in dijet, $WW$ and $t\bar t$
    production processes.
    The distributions are shown for Poisson 
    distributed pileup (with an average of $30$ pileup events) and
    the averages are shown as a function of the number of pileup
    events, $n_\text{PU}$.
    The shapes are calculated for jets with $p_t > 500\GeV$ (the cut
    is applied before adding pileup, as are the cuts on the jet mass
    $m_J$ and subjettiness ratio $\tau_{21}$ where relevant).  }
  \label{fig:shapes}
\end{figure*}
One should be aware that observables constructed from ratios of
shapes, such as $\tau_{n,n-1}$ and planar flow, are not infrared and
collinear (IRC) safe for generic jets.
In particular Pf and $\tau_{21}$ are IRC safe only when applied to jets
with a structure of at least two hard prongs, usually guaranteed by
requiring the jets to have significant mass;
$\tau_{32}$ requires a hard three-pronged structure,\footnote{%
  Consider a jet consisting instead of just two hard particles with
  $p_t = 1000\GeV$, with $\phi=0,0.5$ and two further soft particles
  with $p_t = \epsilon$, at $\phi=0.05,0.1$, all particles having
  $y=0$.
  It is straightforward to see that $\tau_{32}$ is finite and
  independent of $\epsilon$ for $\epsilon \to 0$, which results in an
  infinite leading-order perturbative distribution for $\tau_{32}$.
} a condition not imposed in
previous work, and that we will apply here through a cut on
$\tau_{21}$.

For the angularities and EEC moments we have verified that
the first two numerically-obtained derivatives agree with analytical
calculations in the case of a jet consisting of a single hard
particle.
For variables like $\tau_N$ that involve a partition of a
jet, one subtlety is that the partitioning can change as
the ghost momenta are varied to evaluate the numerical derivative.
The resulting discontinuities (or non-smoothness) in the observable's
value would then result in nonsensical estimates of the derivatives.
We find no such issue in our numerical method to evaluate the
derivatives, but were it to arise, one could choose to force a
fixed partitioning.

To test the method in simulated events with pileup, we use
Pythia~8.165, tune 4C~\cite{Sjostrand:2006za,Sjostrand:2007gs}.
We consider 3 hard event samples: dijet, $WW$ and $t\bar t$
production, with hadronic $W$ decays, all with underlying event (UE)
turned off (were it turned on, the subtraction procedure would remove
it too).
We use anti-$k_t$ jets~\cite{Cacciari:2008gp} with $R=0.7$, taking
only those with $p_t > 500\GeV$ (before addition of pileup).
All jet-finding is performed with FastJet~3.0~\cite{FastJet}.
The determination of $\rho$ and $\rho_m$ for each event 
follows the area--median approach~\cite{Cacciari:2007fd}:
the event is broken into patches and in each patch one evaluates
$p_{t,\text{patch}} = \sum_{i\in\text{patch}} p_{t,i}$, as well as
$m_{\delta,\text{patch}} = \sum_{i\in\text{patch}}
\big(\sqrt{\smash[b]{m^2_{i} + p_{t,i}^2}} - p_{ti}\big)$, where the
sum runs over particles $i$ in the patch.  Then $\rho$ and $\rho_m$ are
given by
\begin{equation}
  \label{eq:rho-rhom-median}
  \rho = \mathop{\text{median}}_\text{patches}
  \left\{\frac{p_{t,\text{patch}}}{A_\text{patch}}\right\}
  \,,\quad
  \rho_m = \mathop{\text{median}}_\text{patches}
  \left\{\frac{m_{\delta,\text{patch}}}{A_\text{patch}}\right\},
\end{equation}
where $A$ is the area of each patch.
To obtain the patches we cluster the event with the $k_t$ algorithm
with $R=0.4$.
For non-zero $\rho_m$ the formula for correcting a jet's 4-momentum
is
\begin{equation}
  \label{eq:4vector-subtraction-with-rhom}
  p^\mu_\text{jet,sub} = 
  p^\mu_\text{jet}  
  - [\rho A^x_\text{jet}, \,
     \rho A^y_\text{jet}, \,
     (\rho+\rho_m) A^z_\text{jet}, \,
     (\rho+\rho_m) A^E_\text{jet}
     ]\,,
\end{equation}
with the area $4$-vector, $A^\mu$, as defined in~\cite{Cacciari:2008gn}.

\begin{figure*}[ht]
  \includegraphics[angle=270,width=0.32\textwidth]{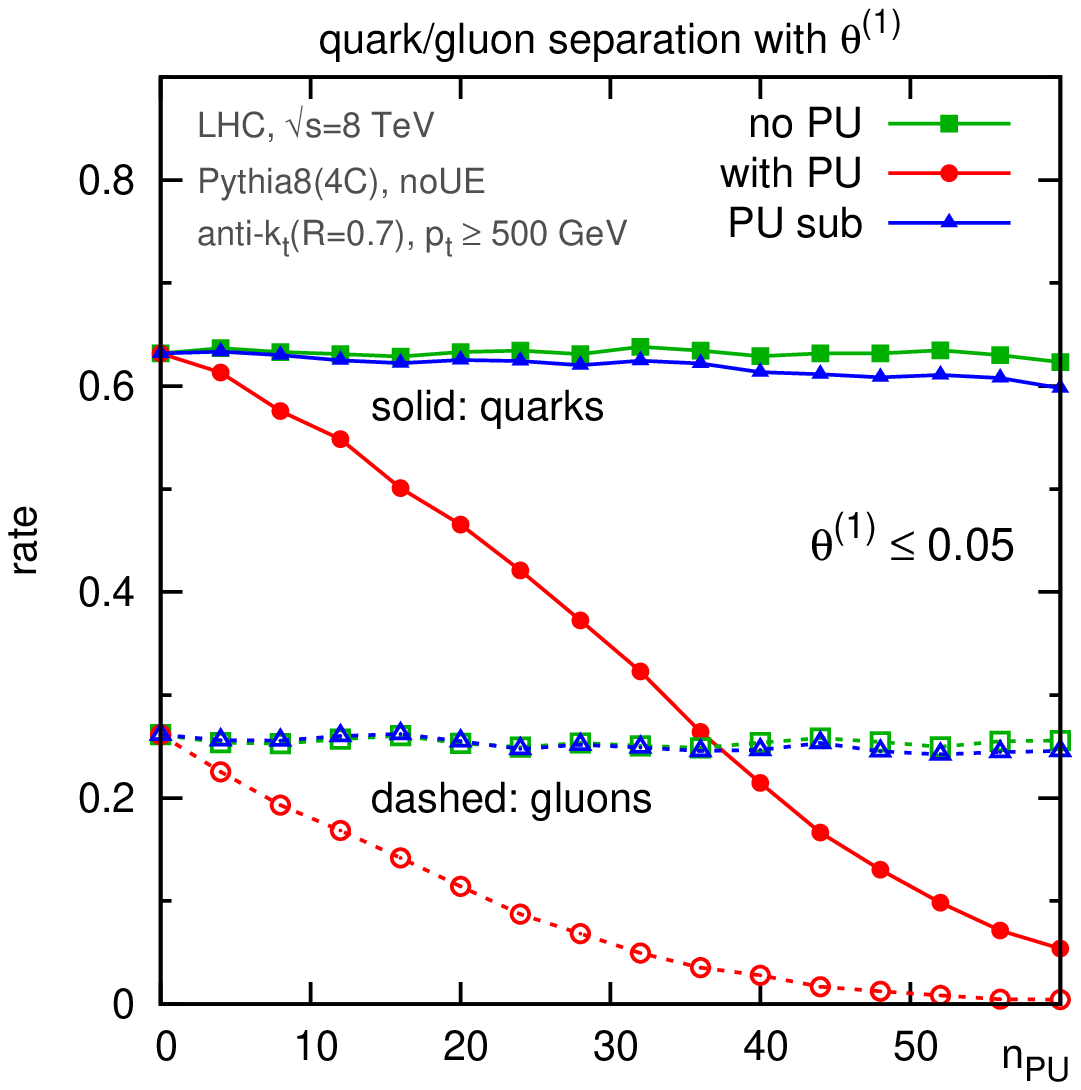}\hfill
  \includegraphics[angle=270,width=0.32\textwidth]{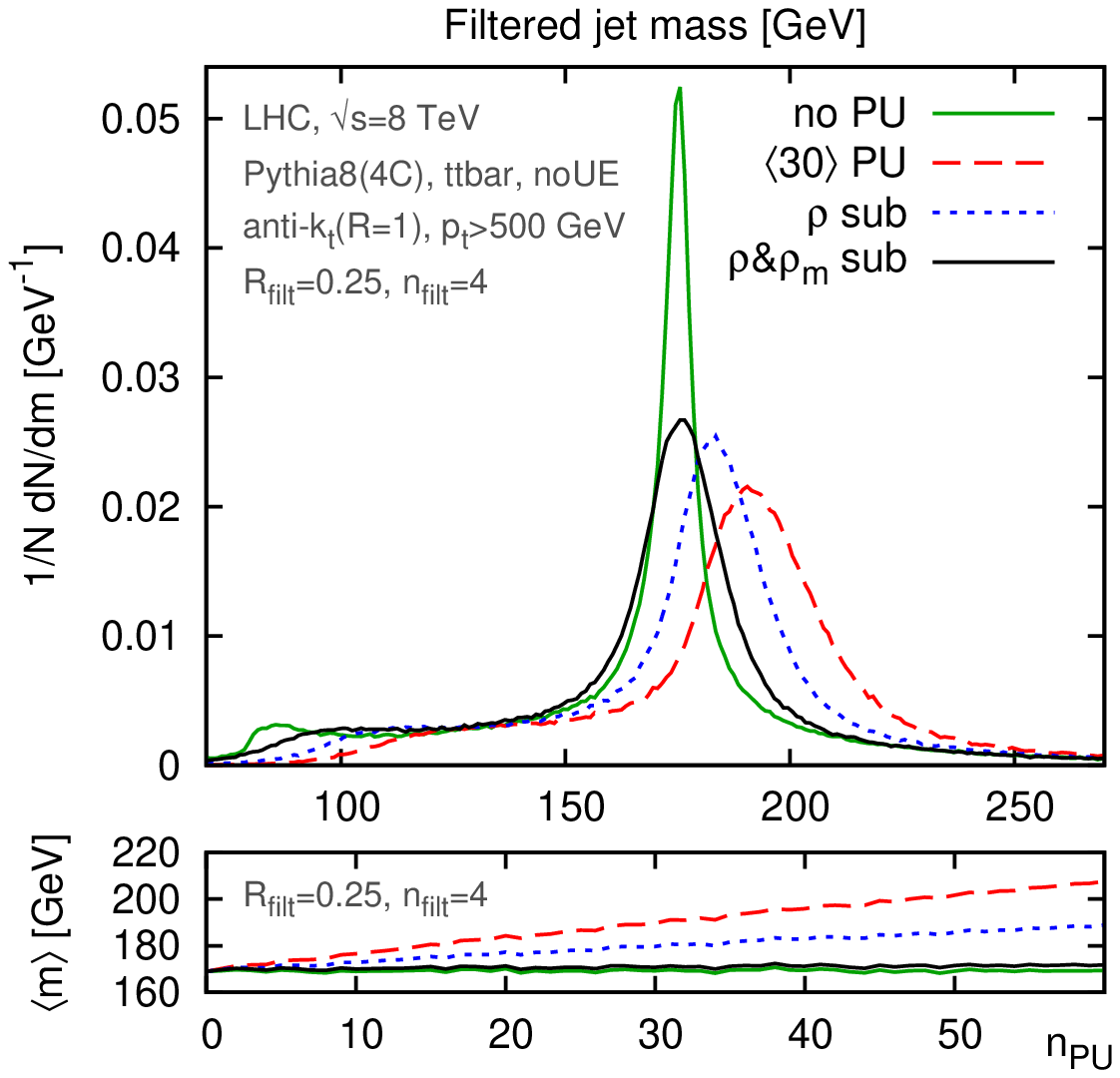}\hfill
  \includegraphics[angle=270,width=0.32\textwidth]{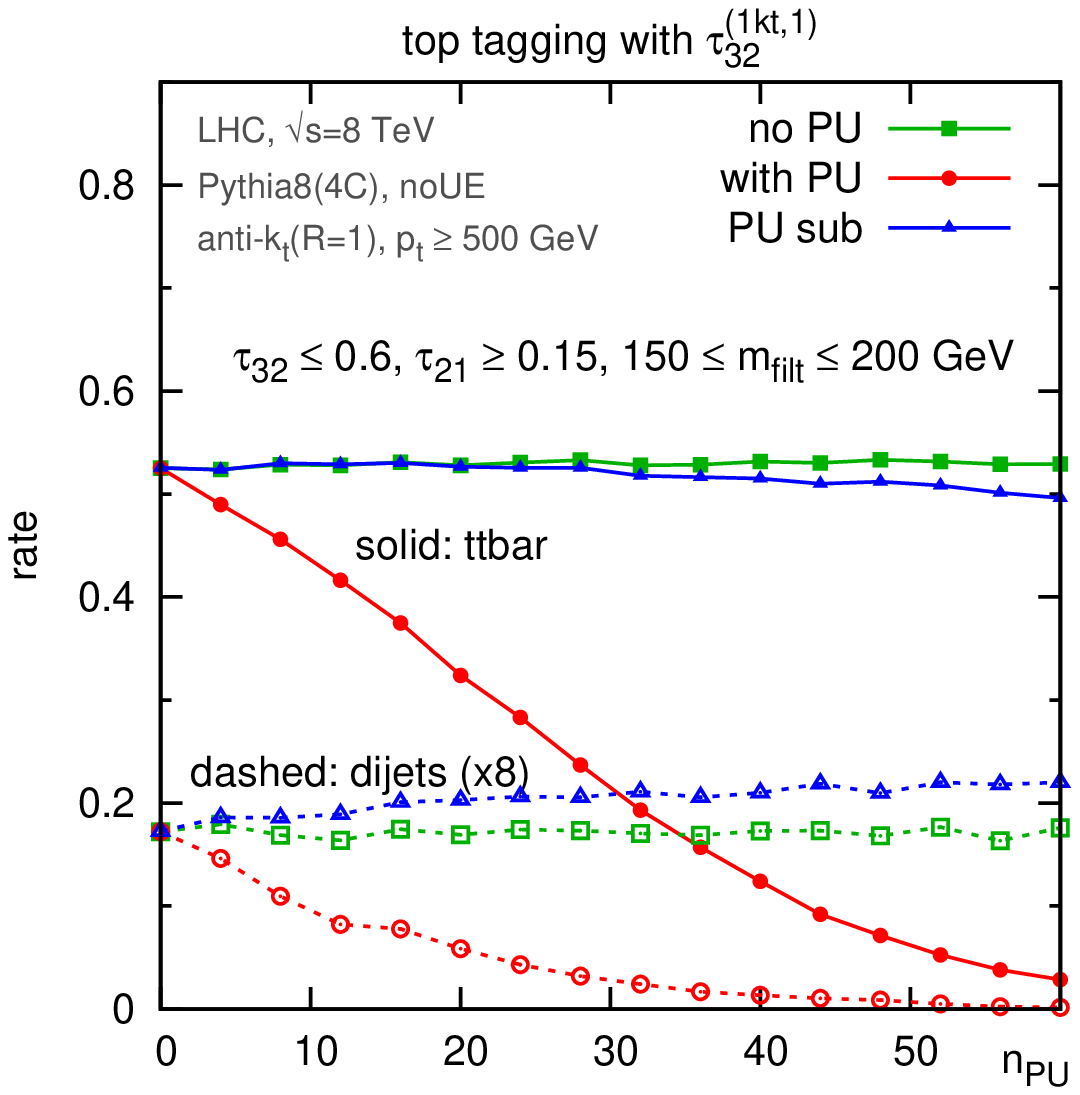}
  \caption{Left: rate for tagging quark and gluon jets using a fixed cut on
    the jet width, shown as a function of the number of pileup
    vertices.
    Middle: filtered jet-mass distribution 
    for fat jets in $t\bar t$ events, showing the impact of 
    the $\rho$ and $\rho_m$ components of the subtraction.
    Right: tagging rate of an $N$-subjettiness top tagger for $t\bar
    t$ signal and dijet background as a function of the number of
    pileup vertices.
    All cuts are applied after addition (and possible subtraction) of
    pileup. Subtraction acts on $\tau_1$, $\tau_2$ and $\tau_3$
    individually. See text for further details. }
  \label{fig:qg}
\end{figure*}

We have 17 observables and 3 event samples.
Fig.~\ref{fig:shapes} gives a representative subset of the resulting
$51$ distributions, showing in each case the distribution (and
average) for the shape without pileup (solid green line), the result
with pileup (dashed line) and the impact of subtracting first and
second derivatives (dotted and solid black lines respectively).
The plots for the distributions have been generated using a Poisson
distribution of pileup events with an average of 30 events (our count
includes diffractive and elastic events, and the analysis uses all
particles from the event generator, leading to $\rho \simeq
770\,\text{MeV}$ and $\rho_m \simeq 125\,\text{MeV}$ per pileup event
at central rapidities).

For nearly all the jet shapes, the pileup has a substantial impact,
shifting the average values by up to $50-100\%$ (as compared to a $5-10\%$
effect on the jet $p_t$).
The subtraction performs adequately: the averaged
subtracted results for the shapes usually return very close to their
original values, with the second derivative playing a small but
sometimes relevant role.
For the distributions, tails of the distributions are generally well
recovered; however intrajet pileup fluctuations cause sharp peaks to
be somewhat broadened. These cannot be corrected for without applying
some form of noise reduction, which would however also tend to
introduce a bias.
Of the 51 combinations of observable and process that we examined,
most were of similar quality to those illustrated in
Fig.~\ref{fig:shapes}, with the broadening of narrow peaks found to be
more extreme for larger $\beta$ values.
The one case where the subtraction procedure failed was the
planar flow for (hadronic) $WW$ events:
here the impact of pileup is dramatic, transforming a peak near
the lower boundary of the shape's range, $\text{Pf}=0$, into a peak
near its upper boundary, $\text{Pf}=1$ (bottom-right plot of
Fig.~\ref{fig:shapes}).
This is an example where one cannot view the pileup as simply
``perturbing'' the jet shape, in part because of intrinsic large
non-linearities in the shape's behaviour; with our particular set of
$p_t$ cuts and jet definition, the use of the small-$\rho$ expansion
of Eq.~(\ref{eq:shape-subtraction-with-rho-Vntotal}) fails to
adequately correct the planar flow for more than about $15$ pileup
events.

Next, we consider the use of the subtraction approach in the context
of quark/gluon discrimination.
In a study of a large number of shapes, Ref.~\cite{Gallicchio:2011xq}
found the jet girth or broadening, $\theta^{(1)}$, to be the most
effective single infrared and collinear safe quark/gluon
discriminator.
Fig.~\ref{fig:qg} (left) shows the fraction of quark and gluon-induced
jets that pass a fixed cut on $\theta^{(1)} \le 0.05$ as a function of
the level of pileup --- pileup radically changes the impact of the cut,
while after subtraction the q/g discrimination returns to its original
behaviour.

Our last test involves top tagging, which we illustrate on $R=1$,
anti-$k_t$ jets using cuts on the ``filtered'' jet mass and on the
$\tau_{32}$ subjettiness ratio.
The filtering selects the 4 hardest $R_\text{filt}=0.25$,
Cambridge/Aachen~\cite{CA} subjets after pileup subtraction.
The distribution of filtered jet mass is shown in Fig.~\ref{fig:qg}
(middle), illustrating that the subtraction mostly recovers the
original distribution and that $\rho_m$ is as important as $\rho$
(specific treatments of hadron masses, e.g. setting them to zero, may
limit the impact of $\rho_m$ in an experimental context).
The tagger itself consists of cuts on $\tau_{32} < 0.6$, $\tau_{21}\ge
0.15 $ and a requirement that the filtered~\cite{Butterworth:2008iy}
jet mass be between $150$ and $200$ GeV.
The rightmost plot of Fig.~\ref{fig:qg} shows the final tagging
efficiencies for hadronic top quarks and for generic dijets as a
function of the number of pileup events.
Pileup has a huge impact on the tagging, but most of the original
performance is restored after subtraction.

To conclude, this letter has shown how most jet shapes can be
straightforwardly corrected for the effects of pileup.
The corrections allow shape-based jet substructure analyses to
continue to perform well even in the presence of up to $60$ pileup
events, notably when combined with the corrections introduced here for
hadron masses in pileup.
This progress will help ensure the viability of a broad range of jet
substructure tools, shape-based and subjet-based, in high-luminosity
LHC running.

The software for the general shape subtraction approach presented here
will be made available as part of the FastJet Contrib
project~\cite{FJContrib}.

\acknowledgements
  We are grateful to David Miller and Jesse Thaler for helpful
  conversations.
  This work was supported by the French Agence Nationale de la
  Recherche, under grants ANR-09-BLAN-0060 and ANR-10-CEXC-009-01 and
  by the EU ITN grant LHCPhenoNet, PITN-GA-2010-264564.

\end{document}